\documentclass[prb,twocolumn,showpacs,preprintnumbers,floatfix]{revtex4} 
\usepackage{amsbsy} \usepackage{graphicx} \usepackage{color}
\usepackage{amsmath} \usepackage{amssymb} \usepackage{amsfonts}
\usepackage{color}

\newlength{\figwidth}
\begin{document}
\title{Compositional uniformity, domain patterning and the mechanism
  underlying nano-chessboard arrays} 
\author{Santiago Gonz\'alez$^{1}$}
\author{J.M. Perez-Mato$^{1}$} \email{wmppemam@ehu.es}
\author{Luis Elcoro$^{1}$} 
\author{Alberto Garc\'{\i}a$^{2}$}
\author{Ray L. Withers$^{3}$} 
\author{L. Bourgeois$^{4}$}

\affiliation{$^1$ Departamento de F\'{\i}sica de la Materia
  Condensada, Facultad de Ciencia y Tecnolog\'{\i}a, Universidad del Pa\'{\i}s
  Vasco (UPV-EHU), Apdo. 644, 48080 Bilbao, Spain.}  
\affiliation{$^2$
  Institut de Ci\`encia de Materials de Barcelona, (ICMAB-CSIC), Campus de la
  UAB, E-08193 Bellaterra, Spain.}  
\affiliation{$^3$
Research School of Chemistry, The Australian National University,
Canberra, A.C.T, Australia}
\affiliation{$^4$
Monash Centre for Electron Microscopy (MCEM), Monash University,
Clayton, Victoria, Australia.}

\date{\today}

\begin{abstract}
We propose that systems exhibiting compositional patterning at the
nanoscale, so far assumed to be due to some kind of ordered phase
segregation, can be understood instead in terms of coherent, single
phase ordering of minority motifs, caused by some constrained drive
for uniformity.  The essential features of this type of arrangements
can be reproduced using a superspace construction typical of
uniformity-driven orderings, which only requires the knowledge of the
modulation vectors observed in the diffraction patterns.
The idea is
discussed in terms of a simple two dimensional lattice-gas model that
simulates a binary system in which the dilution of the minority
component is favored. 
This simple model already exhibits a hierarchy of arrangements similar
to the experimentally observed nano-chessboard and nano-diamond
patterns, which are described as occupational modulated structures
with two independent modulation wave vectors and simple step-like
occupation modulation functions.

\end{abstract}
%


\maketitle

\section{Introduction}\label{sec:Intro}

\begin{figure}[h]
\includegraphics[width=\figwidth]{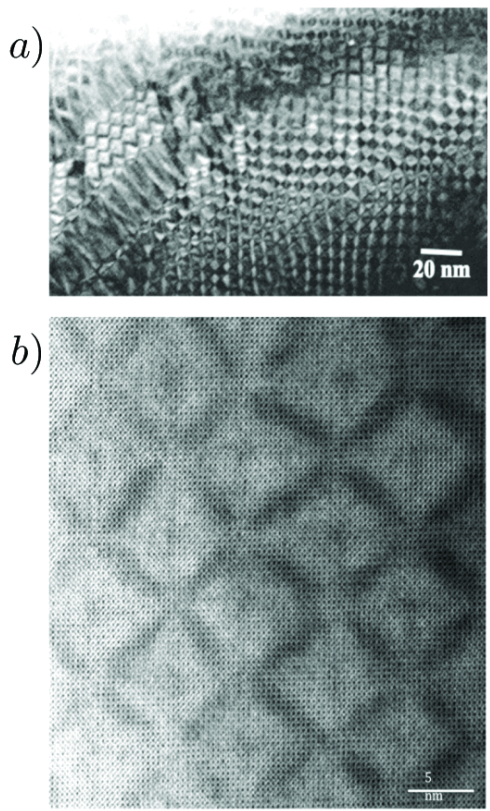}
\caption{ (a) A typical low resolution TEM image of the chessboard
  type nano-structure characteristic of a
  Li$_{1/2-3x}$Nd$_{1/2+x}$TiO$_3$, $x$=0.067, sample. (b) a high
  resolution Bright Field STEM image of this nano-chessboard array.
  The lighter regions are Nd-poor and presumed to correspond to
  regions of stoichiometry Li$_{1/2}$Nd$_{1/2}$TiO$_3$ while the
  darker boundary regions are Nd-rich and presumed to correspond to
  the stoichiometry Nd$_{2/3}$TiO$_3$, according to Ref.~\onlinecite{guiton}.}
\label{fig:GuitonDavies}
\end{figure}

Domain self-patterning on the nanoscale, particularly if the nanoscale
chemical ordering can be engineered or tuned, is the goal
of much research in the nanoscience and nanotechnology
area. Potential applications include the use of the surfaces of such
materials as templates for the assembly of molecular monolayers, for
the reliable synthesis of functional nano-structured materials or for
the structured adsorption of gas species. The recent discovery that
tuneable nanoscale ordering occurs spontaneously in the wide range,
non-stoichiometric Li$_{1/2-3x}$Ln$_{1/2+x}$TiO$_3$, 0.02$\le$ $x$ $\le$
0.12, solid solution~\cite{guiton} as well as in a variety of
other A-site ordered perovskite-related phases
~\cite{garciamartin,licurse2012} is thus of intense current research
interest.  Guiton and Davies~\cite{guiton} demonstrated the
existence of 2D, nanoscale compositional ordering in the
Li$_{1/2-3x}$Nd$_{1/2+x}$TiO$_3$, $0.02\le x \le 0.12$, system via
High Angle Annular Dark Field (HAADF) imaging and interpreted this
result in terms of phase segregation arising from some kind of
``ordered'' spinodal decomposition into 
Li$_{1/2}$Nd$_{1/2}$TiO$_3$ nano-chessboard regions separated by
narrow Nd$_{2/3}$TiO$_3$ boundary regions. As an example,
Fig.~\ref{fig:GuitonDavies} shows the kind of nano-patterns that can
be observed in this compound.

\begin{figure}[h]
\includegraphics[width=\figwidth]{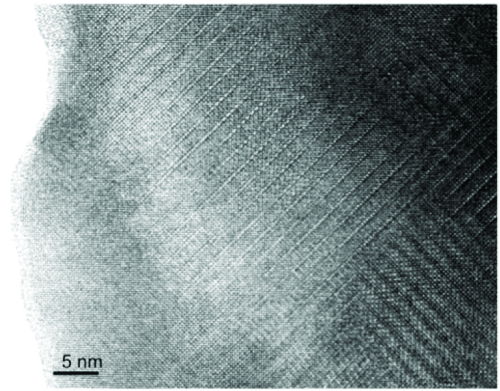}
\caption{ HRTEM image of a striped ordering in
  Bi$_{1-x}$Ca$_x$FeO$_{3-x/2}$, $\sim 0.20 < x < \sim 0.49$, a
  perovskite-related phase viewed along an [001] projection
  direction. Stripes of width 5{\bf a}$_p$ are realized for this particular
  grain (for details see Ref.~\onlinecite{schiemer}).}
\label{fig:Schiemer}
\end{figure}

The 2D apparent phase separation into compositionally distinct areas
described above is strongly reminiscent of the 1D atomic order that
takes place in many compositionally flexible ordered phases. The list
is very extensive, 
but one can cite for instance many binary and/or
ternary alloy systems, ~\cite{schubert,terasaki,watanabe} reduced
rutile structures such as the Ti$_n$O$_{2n-1}$
system,~\cite{bursill1971,terasaki} ReO$_3$ related compounds such as
the MoO$_{3-x}$ system,~\cite{bursill1969} the hexagonal perovskite
related LaTi$_{1-x}$O$_3$ family~\cite{elcoro2000}, the
$(1-x)$Ta$_2$O$_5 +x$WO$_3$ system,~\cite{schmid1995,rae1995} and
others.~\cite{michiue2005} In most of these 1D compositionally ordered
arrangements, the basic driving mechanism is the maximization of the
separation of \emph{minority} structural motifs within a matrix of an
underlying sub-structure. In the case of binary alloy systems, for
example, there is usually a well-defined majority motif such as an
ordered Cu$_3$Au unit which is, as regularly as possible, interleaved
with a minority anti-phase boundary (APB) type structural motif, which
locally has a different composition and leads to a continuously
variable overall composition dependent upon the spatial distribution
of the minority motif. Likewise, in the case of the reduced rutile
Ti$_n$O$_{2n-1}$ system, the majority motif is the rutile type TiO$_2$
structure, while the minority motif is a crystallographic shear plane
(CSP), which again changes the overall composition if ``regularly''
and uniformly arranged.  These compositional orderings are
one-dimensional and often yield
patterns with composition varying in the nanoscale in the form of
stripes (e.g. see
Fig.~\ref{fig:Schiemer}).

In the last decade it has been
shown~\cite{elcoro2000,michiue2005} that this type of one-dimensional
uniformity-driven arrangement can be most efficiently described by
means of step-like atomic occupation modulations with wave vectors and
forms for the modulation functions that can be systematized according
to simple rules using the superspace formalism.
~\cite{janner1980a,janner1980b,wolf1974} Very recently, the main ideas
have been extended to the analysis of the distribution of minority
motifs in \emph{two-dimensional} systems,~\cite{gonzalez2011} showing
that pseudouniform arrangements can originate as the result of a
simple quest for maximal dilution of a minority motif within
lattice-restricted sites.
In wide ranges of composition, the dilution drive was shown to
stabilize 1D stripe arrangements oblique to the underlying lattice,
following predictable rules based on the alternation of sub-units of
different compositions on the nanoscale, but still keeping an overall
identity as one single phase. The stripe arrangements were again shown
to be easily describable as occupationally modulated structures,
needing only a single modulation vector (as befits their intrinsic
one-dimensionality) that can be determined unambiguously from the
diffraction diagram of the system.

In this paper we show that these findings can be extended to the
case of intrinsically two-dimensional patterns, which need \emph{two}
modulation vectors for their superspace description. These 2D pseudouniform
arrangements appear in certain ranges of composition, and in some
cases exhibit nano-chessboard-like features similar to those observed
in the compounds mentioned at the beginning of this section. These
patterns, resulting again from a basic drive to uniformity or maximal
dilution within a lattice, involve a coherent arrangement of
structural motifs, which is rather different from the idea of ordered spinodal
decomposition or phase segregation that has been previously proposed.

The paper is organized as follows. In Sec.~\ref{sec:1Dstripe} we
review the basic ideas and results on pseudouniform stripe orderings
in a simple two-dimensional $A_{1-x}B_x$ binary model presented
recently (Ref.~\onlinecite{gonzalez2011}). In Sec.~\ref{sec:2D} we
discuss the features and superspace description of a different family
of pseudouniform $A_{1-x}B_x$ orderings which are intrinsically
two-dimensional, including the determination of their two modulation
vectors and their direct correspondence with the composition of the system.
Finally, in Sec.~\ref{sec:RealSystems} we show how the
above ideas can be useful to describe orderings in real systems.

\section{pseudouniform 1D stripe compositional ordering}
\label{sec:1Dstripe}

\begin{figure}[h]
\begin{center}
\includegraphics[width=\figwidth]{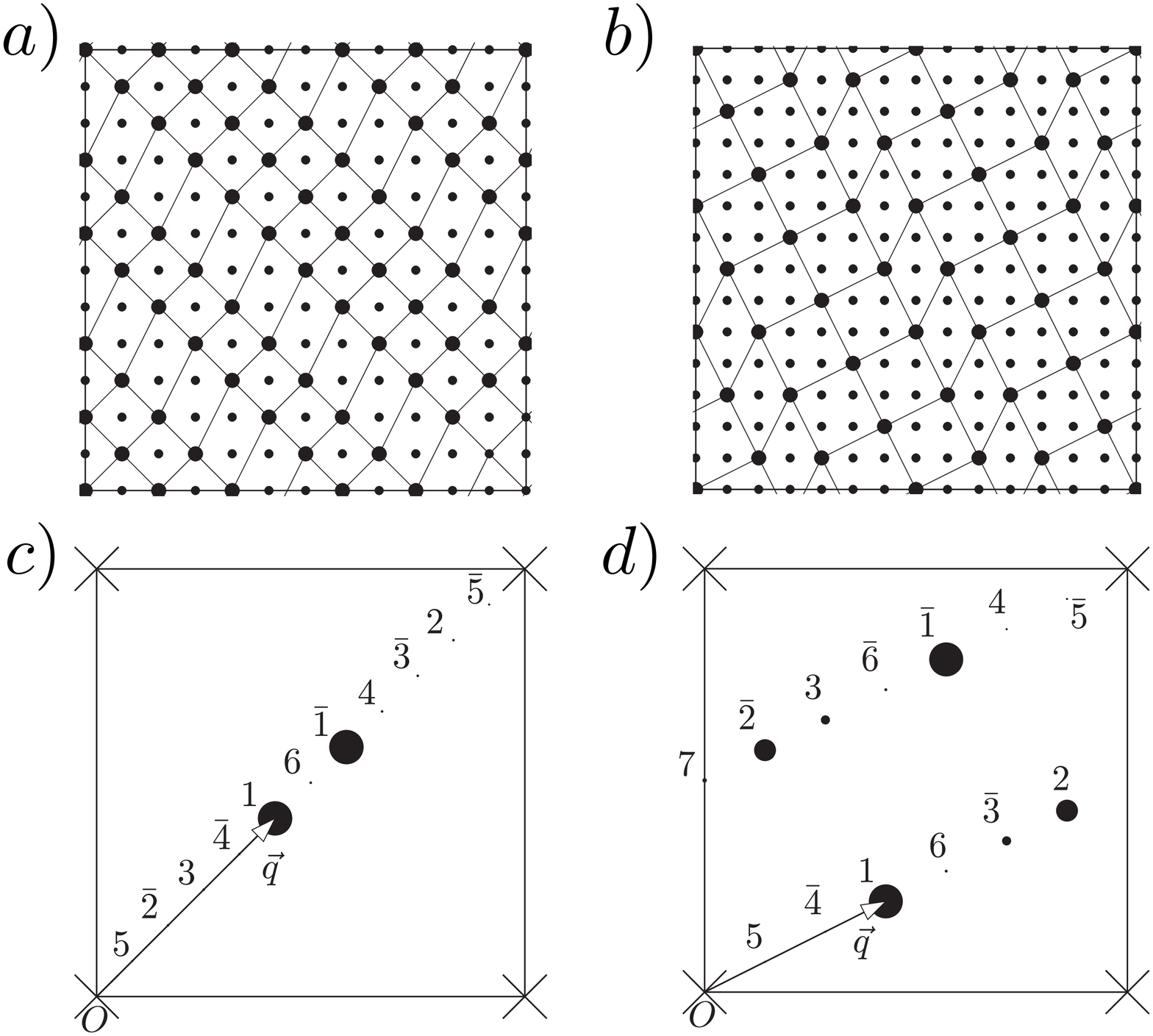}
\end{center}
\caption{Stripe orderings on a square lattice of composition
  $A_{1-x}B_x$ for $x$ = 5/12 (a) and $x$ = 3/14 (b), obtained via
  simulated-annealing optimization using an isotropic Yukawa-type
  repulsive interaction among the $B$
  atoms.~\protect\cite{gonzalez2011} (c) and (d): Fourier spectra
  (discs with radii proportional to the square modulus of the
  geometric structure factor) of the stripe arrangements shown in (a)
  and (b), showing the primary modulation wave vector.  Only
  superstructure reflections are shown. The crosses indicate the
  reciprocal unit cell of the underlying square lattice. The
  superstructure reflections are indexed as satellite reflections
  according to the chosen primary modulation wave vector. }
\label{fig:1Dstripes}
\end{figure}

The basic patterns appearing due to a drive to maximal dilution of
minority motifs, abstracted from any other effects present in real
systems, can be studied using a simple model of a square lattice with
binary composition $A_{1-x}B_x$, in which the $B$ ``atoms'' represent
the minority motifs. In Ref.~\onlinecite{gonzalez2011} the dilution
drive was originally mimicked by isotropic Yukawa-type repulsive
interactions between the $B$ particles, and a simple lattice-gas
simulated-annealing method was used to explore and generate
pseudouniform patterns by minimizing the repulsion energy.  For most
compositions, stripe orderings of the type shown on
Fig.~\ref{fig:1Dstripes} were found to optimize the lattice-ordered
dilution of the $B$ atoms. The stripes are formed by the concatenation
of \emph{tiles}, parallelepipeds delineated by $B$ atoms, with a fixed
composition and orientation within a stripe. For example, in
Figure~\ref{fig:1Dstripes}(a), the $B$ atoms (the larger black dots)
are present in the proportion $x$=5/12 and they achieve an ordered
distribution as close as possible to uniformity by forming a sequence
of $B$-rich and $B$-poor stripes of tiles with composition 1/2 and 1/3
respectively (the smaller, square tile contains one $A$ and one $B$
atom, while the larger tile contains one $B$ and two $A$ atoms).  As
the overall $B$ fraction $x$=5/12 is closer to 1/2 than to 1/3, the
1/2 stripe regions are dominant (by a ratio 3:2), and the 1/3 stripes
have minimal width and are isolated, as if they could be considered
minority motifs themselves, subject to a dilution drive.
The fact that the stripe sequence and makeup follow strict rules as a
function of overall composition was one of the main results of
Ref.~\onlinecite{gonzalez2011}. The rules are based on the
decomposition of the overall $B$ fraction $x$ in irreducible terms
according to the Farey tree construction~\cite{Farey,gonzalez2011}:
for example, $x$=5/12 can be decomposed as 1/2 $\oplus$ 1/2 $\oplus$
1/3 $\oplus$ 1/2 $\oplus$ 1/3, where the $\oplus$ notation indicates a
simultaneous addition of numerator and denominator in the fractions
(e.g. $n_1/m_1 \oplus n_2/m_2$ = $(n_1+n_2)/(m_1+m_2)$, 
which physically corresponds to the concatenation of two
subsets of motifs of concentrations $x_1$=$n_1/m_1$ and
$x_2$=$n_2/m_2$) This leads to the so-called
uniform sequence [22323], which compactly indicates the composition
and arrangement of the basic tiles in a pseudouniform stripe
configuration.

Figure~\ref{fig:1Dstripes}(b) shows an analogous stripe ordering for
$x$=3/14, where stripes with composition 1/5 and 1/4 are arranged
according to the uniform sequence [554]. The stripe pseudouniform
sequence corresponding to any other composition can be derived
using the Farey tree construction. In most cases the arrangement
involves irreducible tiles of compositions $1/n$ and $1/(n+1)$
bracketing the overall $B$ fraction $x$.~\cite{gonzalez2011}. These
tiles can be seen as encoding at a local level the drive for dilution,
and their overall arrangement according to the rules results in a
pseudouniform intergrowth.
As discussed in Ref.~\onlinecite{gonzalez2011}, the details of the
repulsive model potential influence the details of the stability of
different arrangements, so a purely repulsive criterion does not
univocally lead to a single solution in two dimensions. Hence our consistent
use of the term \emph{pseudouniform} to refer to the these orderings.
Fine differences in stability are in practice irrelevant
when one takes into account that the drive for dilution is a
convenient but rough simplification of the state of affairs in real
systems. This suggests that a fruitful approach to the problem of
cataloging pseudouniform orderings in two dimensions is to abstract
the basic ingredients, which are the irreducible, locally uniform,
tiles, and their juxtaposition into intergrowth arrangements following
appropriate rules.

Another key result of Ref.~\onlinecite{gonzalez2011} is that the
arrangements of the type shown in Figure~\ref{fig:1Dstripes} can be
considered as compositionally modulated structures with a single
composition-dependent primary modulation wave vector (a 1D
modulation).~\cite{gonzalez2011} The Fourier spectra (square modulus
of the geometric structure factor) of these two examples are shown in
Figure~\ref{fig:1Dstripes}(c) and~\ref{fig:1Dstripes}(d). One can see
that the wave vectors (5/12)(1,1) and (3/14)(2,1), respectively, which
are approximately orthogonal to the stripe direction, determine the
location of the most intense \emph{satellite} reflections stemming
from the ordering.  Within the
superspace formalism originally developed to describe incommensurately
modulated structures,~\cite{janssen2007,janssen1992,smaalen2007} these
pseudouniform orderings can be described as long-period but commensurate structures
with a simple step-like occupational modulation, with values either
``atom $A$'' or ``atom $B$'' in the relevant proportion, and using
those vectors as primary modulation wave vectors. More specifically, a
pseudouniform striped ordering is fully defined by an occupational
modulation function $f(x_4)$ of period 1 with value ``atom $B$'' in
the interval $~–-x/2 < x_4 < x/2$ and value ``atom $A$'' in the
interval $x/2 < x_4 < 1-x/2$ and a modulation vector {\bf q} that
depends on $x$ according to definite rules,~\cite{gonzalez2011} such that
the kind of atom occupying a site {\bf m} of the lattice is determined
by the value of this function at $x_4$={\bf q}$\cdot${\bf m}. The
continuous variable $x_4$ of the modulation function can be identified
with the coordinate along the \emph{internal space} used in the
superspace formalism.
Thus, when described by means of occupational modulations,
pseudouniform orderings have their modulation functions reduced to
simple consecutive $A$- and $B$-valued intervals (\emph{atomic
  domains}) defined along the internal space in the superspace
formalism.~\cite{gonzalez2011}

\section{Pseudouniform 2D compositional ordering in a prototype $A_{1-x}B_x$
  model}
{\label{sec:2D}}

\subsection{The snub-square ordering}
{\label{sec:Snub}}

In some circumstances more complex, essentially two-dimensional,
ordered configurations in the $A_{1-x}B_x$ system can become
competitive and prevail (in the sense of maximal dilution) over the 1D
striped arrangements studied in Ref.~\onlinecite{gonzalez2011}. For
the case of a square lattice and an ideal isotropic dilution drive
this happens for compositions within the interval 1/5$<x<$1/4. This
composition interval is special because the basic tiles corresponding
to the compositions $x$=1/4 and $x$=1/5 can be juxtaposed with
different orientations (the two sides of the two tiles have the same
length), forming two-dimensional patterns (tilings). These 2D patterns
were already considered in Ref.~\onlinecite{watson} in a simple
context. Here we present a more complete picture of this family of
orderings, their generating principles, and their description as
modulated structures.

\begin{figure}[h]
\begin{center}
\includegraphics[width=\figwidth]{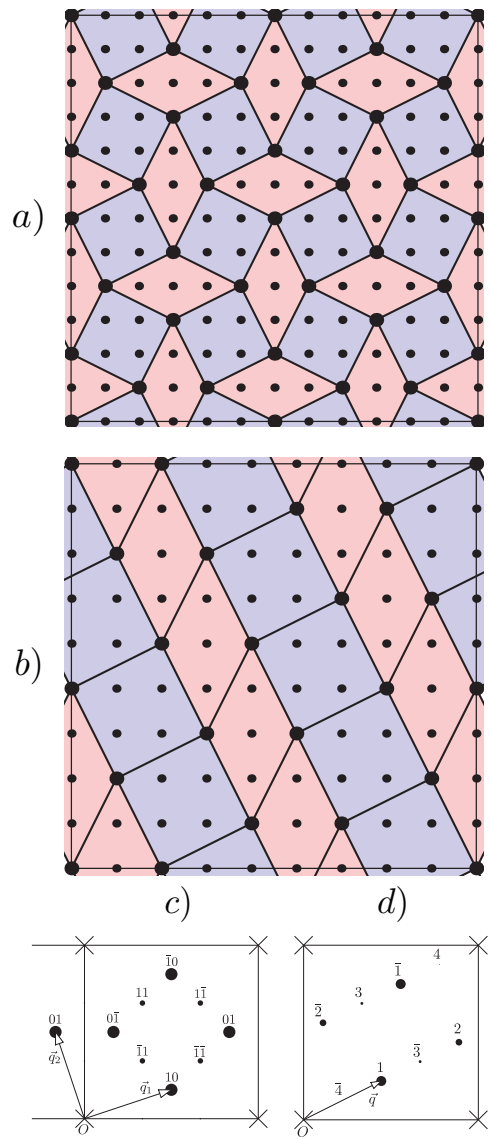}
\end{center}
\caption{ Uniformity-driven orderings in a square lattice of
  composition $A_{1-x}B_x$ for $x$=2/9, obtained with simulations based
  on an isotropic Yukawa-type repulsive interaction among the minority
  $B$ atoms, for 12$\times$12 (a) and 9$\times$9(b)
  simulation boxes, and their corresponding Fourier spectra ((c) and
    (d)). The primary modulation wave vectors, one for the stripe
  arrangement (b) and two for the 2D tiling (a), are indicated in the
  corresponding Fourier diagram. The crosses in (c) and (d) indicate
  the reciprocal unit cell of the underlying square lattice. The
  superstructure reflections are indexed as satellites using the
  primary wave vector(s).}
\label{fig:Snub29}
\end{figure}

Simulations for the repulsive lattice-gas model
mentioned in the previous section indeed confirm the apparition of
intrinsically 2D ordered patterns in this composition
range. Fig.~\ref{fig:Snub29} shows the ground states obtained for a
composition $x$=2/9 for two different sizes of the simulation
supercell (subject to periodic boundary conditions). The pseudouniform
[45] stripe arrangement, in agreement with the rules explained in the
previous section, was obtained for a minimal 9$\times$9 supercell
(Figure~\ref{fig:Snub29}(b)). But for a 12$\times$12 supercell, a
completely different ordering pattern is stabilized
(Figure~\ref{fig:Snub29}(a)).  It has the form of a so-called
snub-square tiling~\cite{grunbaum} (with a slight modification: the
rhombic 1/4 tile can be considered formed by two triangles that are
isosceles instead of being equilateral as in the canonical snub-square
tessellation). This ordering pattern (henceforth SSQ ordering)
prevails in the simulations over the stripe arrangement if the
boundary conditions are compatible with both types of orderings.

A comparison of the distribution of $B$-$B$ distances points to the
main reason for the prevalence of the SSQ ordering over the striped
one. As pointed out by Watson,~\cite{watson} both
have the same distribution of $B$-$B$ interatomic distances up to
$\sqrt{10}$ (in cell parameter units of the underlying square
lattice), but the SSQ ordering avoids the next distance $\sqrt{13}$
associated with neighbours separated by a vector (3,2) or (2,3), which
is present in the striped arrangement.

Figure~\ref{fig:Snub29}(c) and~\ref{fig:Snub29}(d) presents sketches
of the Fourier spectra of the SSQ ordering and of the competing
striped arrangement. They clearly show the primary modulation wave
vector(s) that can be identified in each case. The superstructure
reflections of the striped arrangement can be indexed as linearly
arranged satellites, with the strongest one being given by the
modulation wave vector {\bf q}= (2/9)(2,1), in accordance with the
rules established in Ref.~\onlinecite{gonzalez2011}. The diffraction
pattern of the SSQ ordering requires instead two primary wave vectors
for a simple indexation of the satellite reflections, and the
strongest satellites indicate the most obvious choice: {\bf
  q}$_1$=(1/2,1/6) and {\bf q}$_2$=(--1/6,1/2).

The relevance of the SSQ-type arrangements as 2D orderings that could
maximize somehow the uniformity of the distribution of a minority
component for some specific compositions was pointed out in
Ref.~\onlinecite{watson}. We stress here that these configurations are
indeed observed as ground states in simulations of lattice gas models
with repulsive interactions that mimick an isotropic drive to
uniformity. In the following we will show their hierarchical structure
when they are interpreted in terms of uniform sequences, and how they
can produce nano-chessboard and nano-diamond patterns. In this
framework, it will be shown that this type of arrangements, when
described as compositional modulated structures, have some simple
basic common ingredients which can be generalized and applied to
explain and rationalize orderings in real systems generated at least
in part by a drive to maximal uniformity, including ordered patterns
in the nano scale.

\subsection{Generalization of the snub-square ordering for any
  composition between 1/4 and 1/5.}

\begin{figure}[h]
\begin{center}
\includegraphics[width=\figwidth]{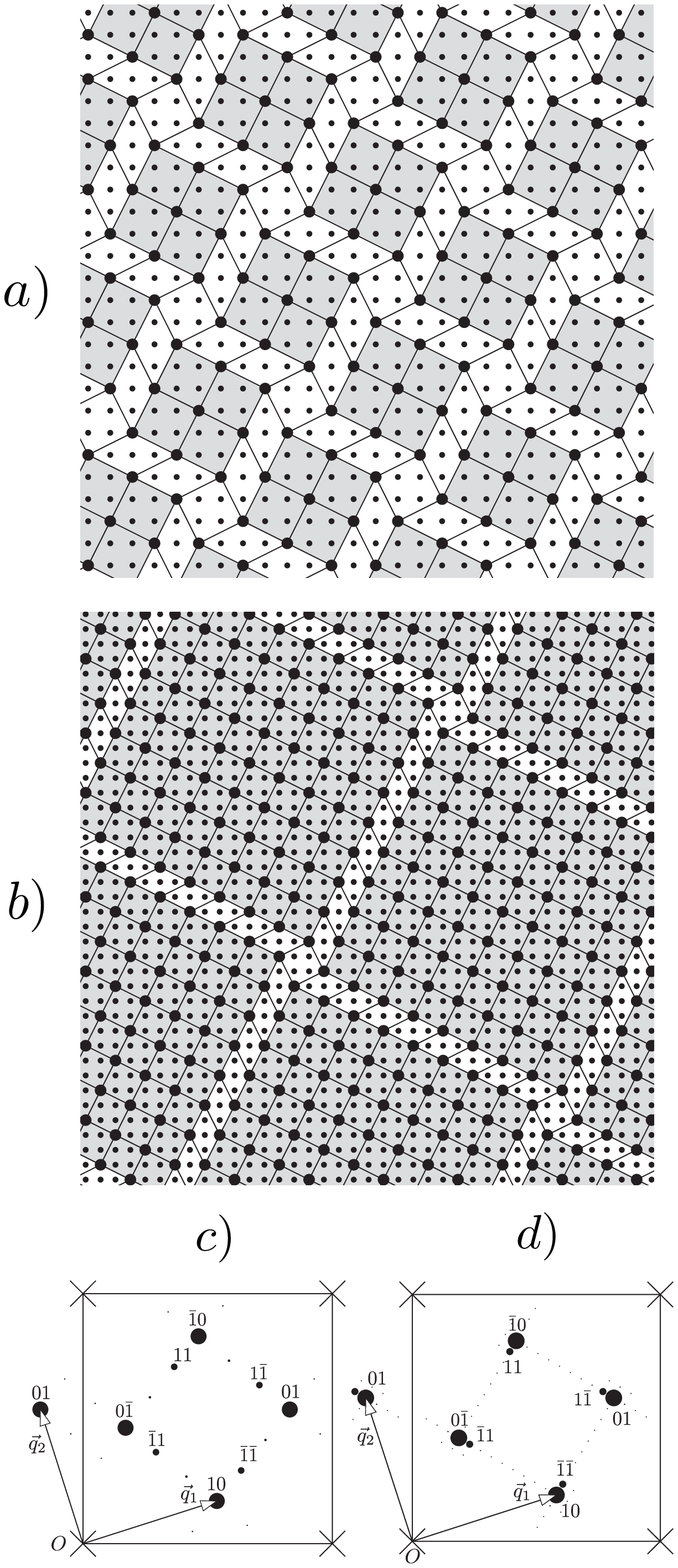}
\end{center}
\caption{Generalized
  snub square orderings in a square lattice of composition
  $A_{1-x}B_x$ for (a) $x$=9/41=0.219512 ($\alpha$=1/3) and (b)
  $x$=81/389=0.20826 ($\alpha$=1/9) and their corresponding Fourier
  spectra ((c) and (d)). The crosses in (c) and (d) indicate the
  reciprocal unit cell of the underlying square lattice. The
  superstructure reflections are indexed as satellites using the
  primary wave vectors that are indicated.}
\label{fig:GenSnub941}
\end{figure}

The SSQ ordering for $x$=2/9 can be seen as a perfect array of
intercrossing ribbons, each one formed by single tiles arranged in a
1/4$\oplus$1/5 sequence. (From now on, to emphasize the actual
arrangement of motifs in a sequence, we will use a slightly modified
notation: \{45\} indicates in this case an alternating sequence of 1/4
and 1/5 tiles, while the [] form will be reserved for pure numerical
Farey term sequences for a specific concentration. In the
stripe-ordering cases both notations are equivalent.) 
This topology, which avoids the distance
$\sqrt{13}$ present in the stripe arrangement, can be generalized to
any other composition in the interval 1/5$<x<$1/4.~\cite{watson} For compositions
different from 2/9, it is not possible to avoid the $\sqrt{13}$
distance completely, but in general the $B$-$B$ distance distribution
will still be favorable when compared to the stripe
ordering.
Following the argument in the previous section, we will discuss these
generalized SSQ orderings in terms of our basic mechanism of
juxtaposing appropriate tiles following certain rules, and we will not
be concerned any more with detailed questions of maximal stability
within a repulsive model. Figure~\ref{fig:GenSnub941}(a) shows an
example of a generalized SSQ ordering with $x<$2/9. It corresponds to
the case $x$=9/41, and is characterized by the crossing of two kinds
of ribbons, one corresponding to the tile sequence \{554\} (shown in
the figure with shaded 1/5 tiles) and the other to the sequence
\{445\}. In each ribbon, the minority tiles appear in a proportion
1/3, and the \{445\} ribbons are also a minority in a proportion
1/3. The minority ratio 1/3 (which we will henceforth denote by
$\alpha$), is thus a key parameter in the construction of the
pattern. In fact, if we take abstract motifs $a$ and $b$, both the
tile sequence in each ribbon and the ribbon arrangement sequence are
instances of the physical sequence \{$aab$\} (uniform sequence [3]), as befits
a concentration $\alpha$=1/3.  By comparison, for the SSQ ordering for
$x$=2/9 in Figure~\ref{fig:Snub29}(a), $\alpha$=1/2, which corresponds
to a simple alternation of 1/4 and 1/5 tiles on each (identical)
ribbon and a simple alternation in the ribbon arrangement (uniform
sequence [2] for $\alpha$=1/2).

The parameter $\alpha$ fully characterizes the generalized SSQ
orderings by determining the hierarchical tiling arrangement, both
at the level of each ribbon, and at the level of ribbon kinds. As it
relates to a minority concentration, we can take without loss of
generality $\alpha\le$1/2, and the general relation between $\alpha$
and $x$ is given by:~\cite{watson}
\begin{equation}
x=\frac{2}{9\pm(1-2\alpha)^2}
\end{equation}
where the plus sign in the denominator corresponds to 1/5$<x<$2/9
(with 1/4 tiles underrepresented), and the minus sign to 2/9$<x<$1/4
(with relative abundance of 1/4 tiles). 

\begin{figure}[h]
\begin{center}
\includegraphics[width=\figwidth]{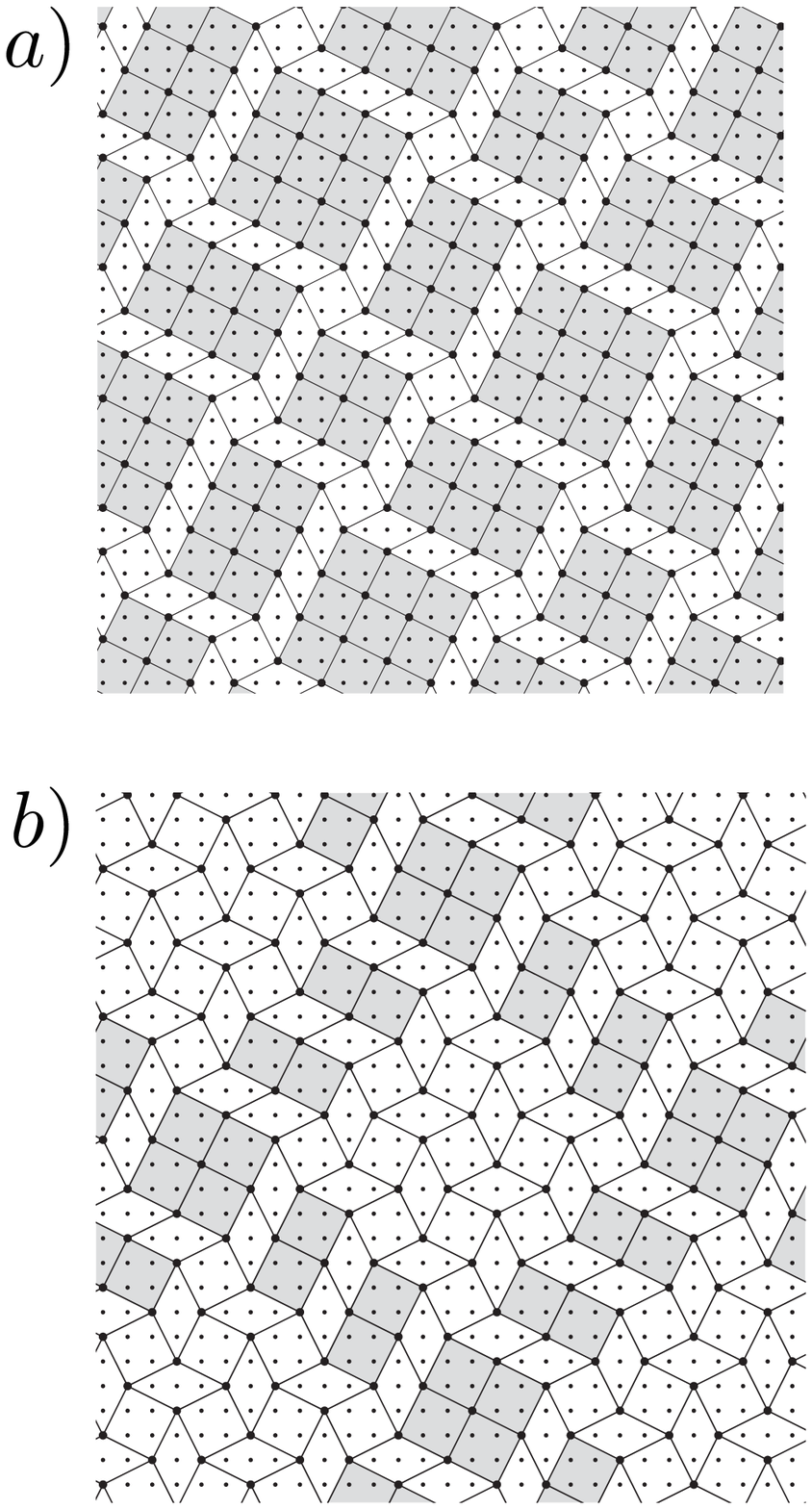}
\end{center}
\caption{ Generalized snub square orderings in a square lattice of
  composition $A_{1-x}B_x$ for (a) $x$=49/225=0.217778 ($\alpha$=2/7)
  and (b) $x$=49/221=0.221719 ($\alpha$=3/7)}
\label{fig:GenSnub225}
\end{figure}

Fig.~\ref{fig:GenSnub941}(b) shows the corresponding arrangement for a
composition much closer to 1/5, $\alpha$=1/9 ($x$=81/389=0.20826). This
ordering optimizes the distribution of $B$-$B$ interatomic distances
towards larger values by forming large squares of composition $x$=1/5,
separated by single ribbons with an inverted 1/9 proportion of 1/5 tiles,
in an effective sequence \{444444445\}. Despite the large size of the square
blocks, the system cannot be considered subject to phase segregation.
On the contrary, the regions with different local composition are
coherently interleaved and ordered as a single phase. Paradoxically,
it is the quest for uniformity that can drive the system into this
type of chessboard orderings.
For $\alpha$ values of type $n/m$, the patterns become more complex,
as the ribbons with a proportion $n/m$ of basic tiles of type
either 1/4 or 1/5 must be ordered according to a 1D non trivial
pseudouniform sequence of the type explained in the previous
section. The presence of 1D uniform sequences (derived from the
Farey-tree construction) along the individual ribbons is in fact the
signature that the ordering is being driven by a uniformity
quest. Figure~\ref{fig:GenSnub225} shows two examples for $\alpha$=2/7
and 3/7.
One can see in this figure that a more complex uniform sequence is
realized in the way the 1/4 and 1/5 basic tiles are arranged along
each oblique ribbon, with either 1/4 or 1/5 tiles as minoritary in a
proportion $\alpha$. For $\alpha$=2/7 (=1/3$\oplus$1/4), the relevant
abstract uniform sequence is \{aabaaab\} ([34]), leading to
interlocking ribbons of the form \{4454445\} and \{5545554\}, which
themselves are in a proportion 2/7 with the same sequencing.

For a rational value of $x$, the corresponding $\alpha$ according to
Eq.~(1) is in most cases irrational, and therefore the generalized
SSQ ordering for that $x$ is incommensurate with respect to
the underlying lattice. The ribbons of basic tiles will follow an
aperiodic uniform sequence corresponding to the irrational value of
$\alpha$, still according to the Farey tree
construction.~\cite{gonzalez2011} Thus, the tendency to uniformity
could conceivably stabilize 2D incommensurate orderings of minority
motifs even if their proportion is a simple rational value. For
instance, the generalized SSQ ordering for $x$=3/14, which competes
with the stripe arrangement shown in Figure~\ref{fig:1Dstripes}(b),
would be an incommensurate ordering with $\alpha$ having an irrational
value close to 2/9.

\begin{figure}[h]
\begin{center}
\includegraphics[width=\figwidth]{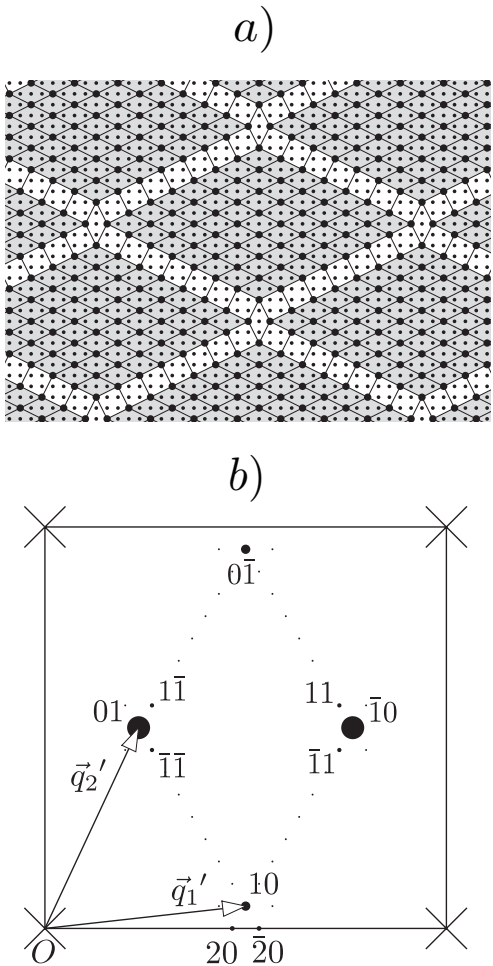}
\end{center}
\caption{ (a) Generalized snub square orderings in a square lattice of
  composition $A_{1-x}B_x$ for $x$=32/135=0.237037 ($\alpha$=1/8,
  $x>$2/9) and (b) sketch of its Fourier spectrum. Superstructure
  reflections are indexed as satellites with primary modulation
  vectors {\bf q'}$_1$ and {\bf q'}$_2$ (which in this case, as
  $x>$2/9, do not follow Eqs.~\ref{eq:q1q2})}
\label{fig:GenSnub135}
\end{figure}

Within the interval 2/9$<x<$1/4 the number of 1/4 tiles is larger
than that of 1/5 tiles, and the ordering patterns follow then the same
rules but with the roles of the two types of basic tiles
interchanged, such that the ribbons having a proportion $\alpha$ of 1/5
tiles become majoritary. As $x$ approaches the composition limit
$x$=1/4, blocks of density 1/4 increase in size, forming patchwork
patterns of diamond shape, as shown in Fig.~\ref{fig:GenSnub135}.

We see that the $\alpha$ parameter acts as an
auxiliary ``minority concentration'', controlling the arrangement
of the 1/4 and 1/5 basic tiles, which in themselves encapsulate a
maximal local uniform ordering. The only other essential ingredient that makes
the SSQs orderings more uniform than the stripe arrangements for the
same composition is the intercrossing feature that tends to eliminate
as much as possible the $\sqrt{13}$ distance.

\subsection{Description of uniformity-driven 2D orderings as
  modulated structures}

Although the generalized SSQ orderings discussed above present a large
variety of arrangements, their diffraction patterns are very
similar. They exhibit a strong hierarchy in the intensity of the
superstructure reflections, characteristic of modulated structures
(see Figures~\ref{fig:Snub29}, \ref{fig:GenSnub941},
and~\ref{fig:GenSnub135}). The strongest superstructure reflections
can be taken as first-order satellites, and define the primary
modulation vectors, while the remaining ones can be indexed as
higher-order satellites. In Section~\ref{sec:Snub} we showed the
primary modulation wave vectors that can be associated with the
$x$=2/9 SSQ ordering. These are particular values among those that can be
obtained for generalized SSQ arrangements, which depend on the
composition according to simple rules of geometrical origin. In
Figure~\ref{fig:GenSnub941}c), for example, the second-order satellite
($\bar{1}$1) lies along the line joining two first-order satellites,
and at distances from them which are fractions $\alpha$ and
$1-\alpha$ of the length of the line. This constraint, together with a
simple symmetry argument (existence of a symmetry plane for $x>$2/9,
and tetragonality for $x<$2/9), is enough to fix the modulation
vectors. To be short, we only describe those corresponding to the
cases with 1/5$<x<$2/9. In this interval, the two modulation wave
vectors are given by the equations:
\begin{eqnarray}
{\bf q}_1= (\frac{1}{2}-\frac{x(1-2\alpha)}{2} , \frac{1}{2}-\frac{3x}{2})
\nonumber\\
{\bf q}_2= (-\frac{1}{2}+\frac{3x}{2} , \frac{1}{2}+\frac{x(1-2\alpha)}{2})
\label{eq:q1q2}
\end{eqnarray}
and are an explicitly tetragonal and right-handed set, best suited for
the construction of the superspace model to follow. (There is an
alternate choice of modulation vectors {\bf k}$_1$={\bf q}$_1$ and
{\bf k}$_2$ =$(1,0) - {\bf q}_2$, with the extra property that
$|${\bf k}$_1\times${\bf k}$_2|$=$x$. Also, these ``natural''
vectors are reciprocal to the real space vectors defining a
``virtual'' monatomic unit cell that best approximates the uniform
motif distribution with density $x$ (in correspondence with the
concept described in
Ref.~\onlinecite{gonzalez2011}).)

%
%

\begin{figure}[h]
\begin{center}
\includegraphics[width=\figwidth]{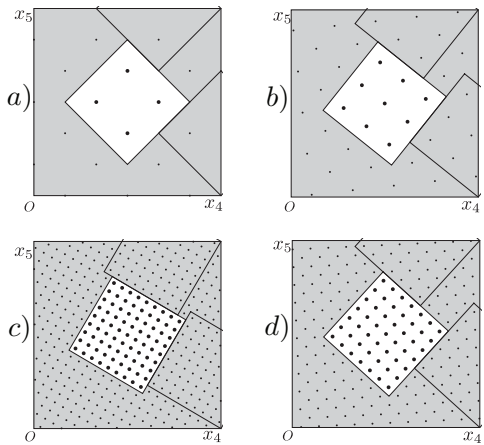}
\end{center}
\caption{ Maps of the occupational step-like modulation functions
  $f(x_4,x_5)$ that describe, taking as modulation wave vectors the
  vectors {\bf q}$_1$ and {\bf q}$_2$ of Eqs.~\protect\ref{eq:q1q2},
  the generalized snub square orderings for $x$=2/9 (a), 9/41 (b),
  81/389 (c) and 49/221 (d), which are shown in
  Figures~\protect\ref{fig:Snub29},\protect\ref{fig:GenSnub941},
  and~\protect\ref{fig:GenSnub225}, respectively.
  The ($x_4$,$x_5$) coordinates of $B$-occupation sites sites are
 indicated with larger dots, and $A$-sites with smaller ones. The $A$
 (dark) and $B$-valued (clear) domains of the simplest function
 consistent with the occupied $A$ and $B$ sites are indicated in each
 case. A single unit cell of the internal space is shown. The
  closeness condition satisfied by the neighboring $B$-valued domains
  in the superspace description is shown by means of continuous lines
  corresponding to their projected borders.}
\label{fig:SuperSpace1}
\end{figure}

%
The SSQ ordering can now be described as a modulated structure with
the modulation vectors of Eqs.~\ref{eq:q1q2} and with an occupational
modulation defined by a function $f(x_4,x_5)$, of period 1 for both
variables, such that the atom occupying a given lattice site {\bf
  m}=($m_1$,$m_2$) is determined by the value (either ``atom $A$'' or
``atom $B$'') of $f(x_4,x_5)$ for $x_4$={\bf q}$_1\cdot${\bf m} and
$x_5$={\bf q}$_2\cdot${\bf m}.
Maps of the function $f(x_4,x_5)$ sampled at the discrete values of
$x_4$ and $x_5$ that correspond to the SSQ orderings considered so far
are shown in Figure~\ref{fig:SuperSpace1}.  
As it happens in the 1D case, the domain with value ``atom $B$'',
corresponding to the minority motifs, forms a single compact/dense
region with an area equal to their proportion $x$ in the lattice. Thus
the minority atoms $B$ aggregate in the internal space of the
occupational modulation function in order to describe arrangements in
which these atoms are maximally scattered in real space.
(As remarked above, for irrational $\alpha$ 
one would obtain from Eqs.~\ref{eq:q1q2} incommensurate wave vectors,
and the sampling of the $A$- and $B$-valued domains would be continuous.)
The $B$ atomic domains have in all cases a form which
avoids the occupation by $B$ atoms of two neighbouring lattice sites.
This restriction is ensured by the
geometrical condition shown in Figure~\ref{fig:SuperSpace1}: The $B$
atomic domains, if translated on the plane $(x_4,x_5)$ by either
({\bf q}$_1\cdot$(1,0),{\bf q}$_2\cdot$(1,0)) or 
({\bf q}$_1\cdot$(0,1),{\bf q}$_2\cdot$(0,1)),
juxtapose with the original one, with no superposition. Within the
superspace formalism this is termed ``closeness condition'',
and is known to be satisfied by the atomic domains describing some
quasicrystals.~\cite{cornier1992,katz1993} The atomic domain borders
can be taken parallel to the basis vectors of the point lattice in
internal space formed by the points
({\bf q}$_1\cdot${\bf m},{\bf q}$_2\cdot${\bf m}), where {\bf m}
represents the direct lattice points.
This, together with the closeness condition and Eqs.~\ref{eq:q1q2},
is sufficient to define the appropiate atomic domain $B$ for any
composition $x$, and to construct with it its corresponding SSQ ordering.

\section{Uniformity-driven orderings in real systems}
\label{sec:RealSystems}

\begin{figure}[h]
\begin{center}
\includegraphics[width=\figwidth]{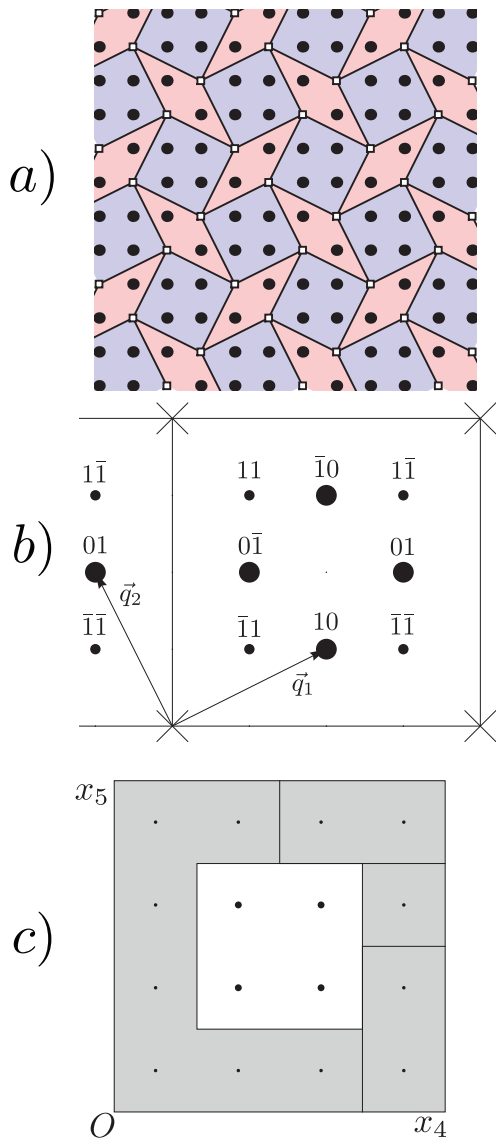}
\end{center}
\caption{ (a) Ordering pattern of the oxygen vacancies reported in the
  layers BO$_{2-\delta}$ of La$_{8-x}$Sr$_x$Cu$_8$O$_{20-\delta}$,
  (La,Sr)$_8$Cu$_8$O$_{18}$, and La$_2$Sr$_6$Cu$_8$O$_{16}$
  (Ref.~\protect\onlinecite{hadermann2005}), interpreted as a
  1/3$\oplus$1/5 tiling in a virtual square lattice in which vacancies
  are present in a proportion $x$=1/4. (b) Diffraction diagram showing
  the modulation vectors {\bf q}$_1$=(1/2,1/4) and {\bf
    q}$_2$=(--1/4,1/2) that index the satellites. (c) Map of the
  modulation function that can be associated with the ordering pattern
  using {\bf q}$_1$ and {\bf q}$_2$ (symbols and extra
  closeness-condition lines as in Figure~\ref{fig:SuperSpace1}).}
\label{fig:OVacancies}
\end{figure}

We have considered above an idealized $A_{1-x}B_x$ model for which the
optimal ordering within a lattice depends only on the maximization of
the dilution of minority motifs. In real systems this factor can be
one among many others. For instance, intrinsic anisotropies of the
underlying crystalline lattice, energy constraints on possible
orderings (for instance, compatibility with rigid unit modes,
chemically forbidden geometries, etc.) can play a fundamental role in
favoring a particular ordering pattern. Nevertheless, if an effective
dilution still retains a main role in these realistic scenarios, some
of the features of the purely uniformity-driven orderings are very
likely to be maintained.
One of the features which are present in the above discussion is the
possibility of constructing the orderings through the juxtaposition
(intergrowth) of basic tiles. These encapsulate at the local level the
drive for dilution, and their combination, as we have seen, can lead
to a variety of patterns, depending on the geometrical flexibility
involved.
Fig.~\ref{fig:OVacancies}(a) shows a sketch of the ordering pattern of
oxygen vacancies observed in layers of the compounds
La$_{8-x}$Sr$_x$Cu$_8$O$_{20-\delta}$, (La,Sr)$_8$Cu$_8$O$_{18}$ and
La$_2$Sr$_6$Cu$_8$O$_{16}$.~\cite{hadermann2005} To our knowledge,
this is the first time that this ordering is interpreted as a tiling,
similar to the SSQ, but with tiles of composition 1/3 and 1/5, so that
the global vacancy density is 1/4. In this case the ordering is not
optimizing a purely repulsive interaction (which would lead for this
composition to a monatomic superlattice of vacancies), but the drive
for dilution is still obviously present.  Considered as an abstract
geometrical and numerical exercise, the combination rules for the 1/3
and 1/5 tiles lead to the same kind of regularities discussed above
for the 1/5, 1/4 tiles, including the appearance of coherent
generalized nano-patches (nano-rhombi in this
case).~\cite{unpublished} Crucially, all the arrangements produced
through the 1/5--1/3 tile combinations are also describable as
occupationally modulated structures, with compact atomic domains in
the superspace construction, and with modulation vectors that can be
read directly from the diffraction diagram. For the case of the
vacancy distribution in Fig.~\ref{fig:OVacancies}(a), the primary
modulation wave vectors can be chosen as {\bf q}$_1$=(1/2,1/4) and
{\bf q}$_2$=$(-1/4,1/2)$, as seen in the diffraction diagram of
Fig.~\ref{fig:OVacancies}(b), and the corresponding modulation
function is shown in Figure~\ref{fig:OVacancies}(c). The four
independent vacancies within the 16 atomic sites form again a single
simple atomic domain fulfilling the closeness condition described
above.

Once again we see that the ordered patterns are describable as
modulated phases with step-like modulation functions, with the
$B$-valued regions of the functions limited to one or a few domains
within the unit-cell of the periodic function, and with the fulfilment
of some kind of closeness condition. Modulated structures with
modulation functions of this type are very efficient for distributing
regularly minority motifs in real space. With all the evidence
presented so far (here and in the previous work over several years
)~\cite{perezmato1999,elcoro2000,boullay2002,darriet2002,michiue2006,izaola2007}
we can say that in a real system the knowledge of the relevant
modulation vectors (extracted from its diffraction diagram) is
sufficient to design, through the rules explained above, a
compositionally modulated model
which corresponds to an atomic ordering with a high degree of
uniformity in the distribution of some specific motif. Thus we can
postulate an atomic distribution model with a specific composition and
a maximal dilution consistent with the observed modulation vectors. If
a drive for dilution is at work, this apriori model should approximate
the experimental arrangement.  This clearly occurs in the example
shown in Fig.~\ref{fig:OVacancies}, and also seems to happen in the
nano-chessboard arrangements observed in the compounds mentioned in
the introduction. As shown below, the basic features of the latter
can be reproduced in a simplified binary system by postulating a
compositionally modulated arrangement with step-like atomic domains that
fulfill the closeness condition, and are consistent with the observed
modulation vectors. This would indicate that the observed arrangements
represent some optimization of the dilution of the atoms within the
underlying perovskite lattice, but subject to some anisotropic
contraints that force the possible modulation vectors.

\begin{figure}[h]
\begin{center}
\includegraphics[width=\figwidth]{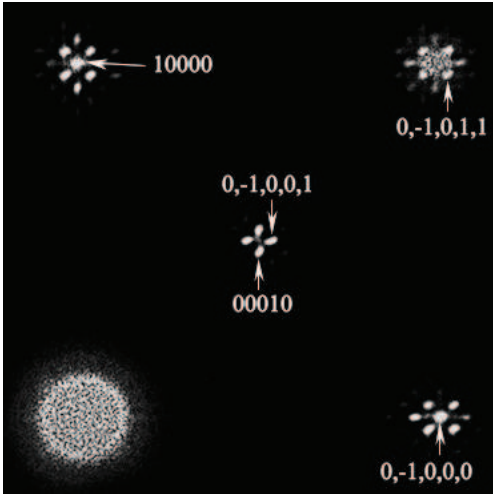}
\end{center}
\caption{[001] zone electron diffraction pattern of the (3+2)-D
  incommensurately modulated, nano chessboard phase of
  Li$_{0.30}$Nd$_{0.567}$TiO$_3$, indexed with respect to the basis
  set M$^*$ = \{ {\bf a}$^*$= {\bf a}$_p^*$, {\bf b}$^*$= {\bf
    b}$_p^*$, {\bf c}$^*$= 1/2{\bf c}$_p^*$, {\bf q}$_1$ =
  1/2{\bf a}$_p^*$ + (1/2--$\epsilon$){\bf b}$_p^*$ , {\bf q}$_2$ =
  (--1/2+$\epsilon$){\bf a}$_p^*$ + 1/2{\bf b}$_p^*$\}, where the
  subscript $p$ stands for parent perovskite, and $\epsilon$ $\sim$
  1/30.}
\label{fig:Ray}
\end{figure}

\begin{figure}[h]
\begin{center}
\includegraphics[width=0.8\figwidth]{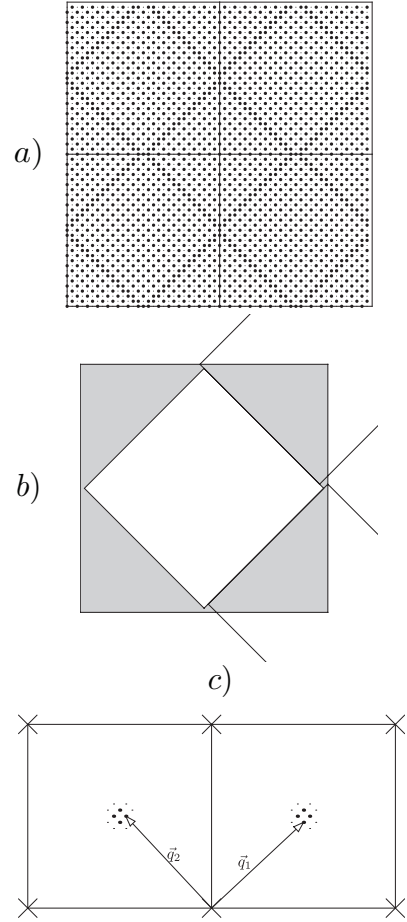}
\end{center}

\caption{ (a) Chessboard nanopattern in a square lattice of
  composition $A_{1-x}B_x$, obtained as a compositional modulated
  structure with modulation wave vectors {\bf
    q}$_1$=(1/2,1/2$-\epsilon$) and {\bf q}$_2$=(--1/2$+\epsilon$,1/2)
  with $\epsilon$=1/30 (i.e. those present in
  Li$_{1/2-3y}$Nd$_{1/2+y}$TiO$_3$), and a step-like $A$/$B$
  occupational modulation function fully determined by the closeness
  condition of the B domains for these modulation vectors. This
  modulation function is shown in (b), where $A$ and $B$-valued domains are
  depicted as dark and clear regions, respectively, within a 2D period
  of the function (i.e. a unit cell of the internal space in the
  superspace construction). The
  closeness condition satisfied by the neighboring $B$-valued domains
  in the superspace description is shown by means of continuous lines
  corresponding to their projected borders.  The corresponding
  geometric diffraction pattern is shown in (c). }
\label{fig:ModelExp}
\end{figure}

Let us consider for instance the compound
Li$_{1/2-3y}$Nd$_{1/2+y}$TiO$_3$
mentioned in the introduction. Figure~\ref{fig:Ray} shows a [001] zone
electron diffraction pattern (EDP) of this material with $y$=0.067 (we
change the letter for the composition variable to avoid confusion with
that employed in our binary toy model). The primary modulation wave
vectors can be identified directly from the EDP diagram as
{\bf q}$_1$=(1/2$-\epsilon$,--1/2) and {\bf
  q}$_2$=(1/2,1/2$-\epsilon$) %
with $\epsilon$ about 1/30.
One cannot pretend to derive from this information a quantitative
structural modelling of this compound, but we can determine for our
simple 2D $A_{1-x}B_x$ binary system a pseudouniform ordering of the B
atoms consistent with these modulation vectors. Using the rules
explained above, we have to consider an occupational modulation
function $f(x_4,x_5)$ fulfilling the closeness condition for these
modulation vectors. This simple step-like occupational modulation
function (see Fig.~\ref{fig:ModelExp}(b)), yields the atom ordering
shown in Fig.~\ref{fig:ModelExp}(a), which vividly recalls the
nano-chessboard arrangements observed in this compound.  The square
patches have composition AB, while the global excess of A atoms are
localized at the interfaces, which are made only of A atoms and are
limited to two unit cells of the underlying lattice. By construction,
the composition of this $A_{1-x}B_x$ nano-chessboard arrangement is
fully determined by the modulation vectors, and is given by the
relative area of the B atomic domain in Fig.~\ref{fig:ModelExp}(b),
which is $x$=1/2$-\epsilon + \epsilon^2$, i.e. $x$=421/900=0.467778.
Obviously this composition is not comparable with that of
Li$_{1/2-3y}$Nd$_{1/2+y}$TiO$_3$. The arrangement of
Fig.~\ref{fig:ModelExp} is far from the real system, not only due to
the reduction to a 2D $A_{1-x}B_x$ arrangement, but also because the
vacancies accompanying the Nd/Li substitution and the expected very
large positional relaxations are ignored.  But nevertheless, this
simple occupational modulation with an intrinsic tendency to
uniformity and consistent with its modulation vectors in a binary
system, is sufficient to reproduce basic features of the real
system. Notice for instance the shift of the $A$ and $B$ occupation
sites in contiguous patches of composition $A_{1/2}B_{1/2}$ of the
chessboard, as proposed in the model for
Li$_{1/2-3y}$Nd$_{1/2+y}$TiO$_3$ of Ref.~\onlinecite{guiton}.
Fig.~\ref{fig:ModelExp}(c) shows the geometric diffraction pattern of
this 2D $A_{1-x}B_x$ nano-chessboard arrangement. It is remarkable
that despite the strong anharmonicity of the occupational modulation
only odd-order satellites are observable. This is consistent with the
Fourier decomposition of the underlying two dimensional step-like
occupational modulation.
In the real system even-order satellites close to the main
reflections are also significant, which is probably due to the strong
displacive modulations in the perovskite framework (mainly tiltings),
induced by the cation ordering. This can be a
plausible explanation, since by just introducing a small displacive
sinusoidal modulation with wave vectors
{\bf q}$_1$+{\bf q}$_2$ and {\bf q}$_1$--{\bf q}$_2$
of the $A$ and $B$ positions in the configuration of
Fig.~\ref{fig:ModelExp}(a), we can produce a diffraction
diagram similar to the experimental one of Figure~\ref{fig:Ray}.

Nano-chessboards of the type reported in
Refs.~\onlinecite{guiton,garciamartin,licurse2012} may therefore
originate in a drive for maximal uniformity in the distribution of a
minority motif.
If this were indeed the mechanism, the width of the interface regions
between the chessboard patches, although clearly wider than in our model due to
structural displacive relaxations, should be independent of the size
of the patches, i.e. of the system composition. A careful check of the
fulfillment or not of this property by the nano-chessboards observed in
real systems would be a key feature for the validation of a 
uniformity-drive mechanism.


\section{Conclusions}

We have shown that a drive towards dilution of minority structural
motifs constrained to an underlying lattice is sufficient to produce
two-dimensional orderings with local composition variations in the
nanoscale.  The dilution can be achieved in practice by the
combination or intergrowth of basic tiles that encapsulate a local
repulsion of minority motifs. This combination principle is able to
generate orderings that are hierarchically structured, adopting for
certain compositions the form of nano-chessboards or diamond-like
patches. In all cases, regions of two different effective compositions
are interleaved coherently, with long-range order.

The mechanism leading to composition patterning from a dilution drive
has already been documented in other systems: layer arrangement in
compositionally flexible layered
compounds,~\cite{perezmato1999,elcoro2000,boullay2002,darriet2002,michiue2006,izaola2007}
and two-dimensional systems exhibiting effectively 1D stripe
patterns.~\cite{gonzalez2011}
Here we find what appears
to be a segregation of phases instead of a pseudouniform
distribution, but these arrangements are in fact near-optimally
uniform by any reasonable criterion, and their underlying structure
(now essentially two-dimensional instead of effectively 1D)
can be described by the same recipe at work in the pseudouniform
systems previously studied: modulation
vectors directly and simply determined from the intensity distribution
in the diffraction diagram, and step-like occupational modulations
satisfying a closeness condition for these modulation vectors.

The description in terms of occupational modulations, best represented
using the tools of the superspace formalism, is thus seen as the
unifying structural principle in a wide variety of systems which exhibit
in some degree a dilution drive. The key idea is that minority motifs
maximally scattered in real space are represented by compact regions
in the internal coordinates of the superspace construction.

We have shown that this powerful principle seems to be at work for two
kinds of relevant experimental systems. In the case of the
distribution of vacancies in La$_{8-x}$Sr$_x$Cu$_8$O$_{20-\delta}$,
(La,Sr)$_8$Cu$_8$O$_{18}$ and
La$_2$Sr$_6$Cu$_8$O$_{16}$,~\cite{hadermann2005} the resulting pattern
can be seen also as a direct example of the tiling rules in real
space. The nano-chessboard arrays, and similar patterns exhibiting
composition patterning, that have been recently observed in
composition flexible systems such as Li$_{1/2-3x}$Nd$_{1/2+x}$TiO$_3$
are strongly reminiscent of the kinds of hierarchical real-space
orderings that appear in uniformity-driven simple models. 
They reproduce basic features of the observed nano-chessboard
arrangements, and their Fourier spectra exhibit the kind of strong
hierarchical satellite structure present in the experimental
diffraction diagrams.
We can thus suggest that the observed compositional changes at the
nanoscale in these compounds have their origin, at least partially, in
a drive for uniformity, maintaining the coherency of a single phase,
and are not, as previously proposed, due to any kind of phase
separation.

\section*{Acknowledgements}

This work has been supported by the Spanish Ministry of Science and
Innovation (projects MAT2008-05839 and FIS2009-12721-C04-O3) and by the
Basque Government (project IT-282-07). Technical and human support
provided by IZO-SGI SGIker (UPV/EHU, MICINN, GV/EJ, ERDF, ESF) is
gratefully acknowledged. RLW acknowledges financial support from the
Australian Research Council in the form of ARC Discovery Grants.


\end{document}